\documentclass[aps,floatfix,twocolumn,floats,superscriptaddress]{revtex4-1}
\usepackage[T1]{fontenc}
\usepackage{times}
\usepackage{graphicx}
\usepackage{amsmath}
\usepackage{amsfonts}
\usepackage{amssymb}
\usepackage{color}
\usepackage{braket}
\usepackage{textcomp}
\usepackage{comment}

\usepackage[normalem]{ulem}  

\newcommand{\ham}{\hat{\mathcal{H}}}

\begin{document}

\title{Reference excitation energies of increasingly large molecules: a QMC study of cyanine dyes}

\author{Alice Cuzzocrea}
\affiliation{MESA+ Institute for Nanotechnology, University of Twente, P.O. Box 217, 7500 AE Enschede, The Netherlands}
\author{Saverio Moroni}
\email{moroni@democritos.it}
\affiliation{CNR-IOM DEMOCRITOS, Istituto Officina dei Materiali, and SISSA Scuola Internazionale Superiore di Studi Avanzati, Via Bonomea 265, I-34136 Trieste, Italy}
\author{Anthony Scemama}
\email{scemama@irsamc.ups-tlse.fr}
\affiliation{Laboratoire de Chimie et Physique Quantiques, Universit\'e de Toulouse, CNRS, UPS, France}
\author{Claudia Filippi}
\email{c.filippi@utwente.nl}
\affiliation{MESA+ Institute for Nanotechnology, University of Twente, P.O. Box 217, 7500 AE Enschede, The Netherlands}

\begin{abstract}
We revisit here the lowest vertical excitations of cyanine dyes using quantum Monte
Carlo and leverage on recent developments to systematically improve
on previous results. In particular, we 
employ a protocol for the construction of compact and accurate multi-determinant Jastrow-Slater
wave functions for multiple states, which we have recently validated on the
excited-state properties of several small prototypical molecules.
Here, we obtain quantum Monte Carlo excitation energies in excellent
agreement with high-level coupled cluster for all the cyanines where the coupled
cluster method is applicable. Furthermore, we push our protocol to longer chains,
demonstrating that quantum Monte Carlo is a viable methodology to establish reference data 
at system sizes which are hard to reach with other high-end approaches of similar accuracy.  Finally, we
determine which ingredients are key to an accurate treatment of these challenging
systems and rationalize why a description of the excitation based on only active $\pi$ orbitals lacks
the desired accuracy for the shorter chains.
\end{abstract}

\maketitle

\section{Introduction}
\label{sec:intro}

Cyanine dyes are a family of charged $\pi$-conjugated molecules which are employed in very diverse
applications ranging from dye-synthesized solar cells to the labeling of bio-molecules~\cite{levitus2011,
shindy2017, oprea2020}.  Their characteristic structure consists of a chain of an odd number of carbons
with two amine groups at the ends. While their photo-physical properties are strongly regulated by the
length of the carbon chain, the lowest bright state of the cyanines always maintains a $\pi\to\pi^*$
character and can be predominantly described as a HOMO to LUMO (HL) transition. Despite the apparent
simplicity of this excitation, its accurate treatment is known to be challenging and, consequently,
cyanine dyes have often been used as model systems to assess the quality of electronic structure
methods for excited
states~\cite{send2011,jacquemin2012,moore2013,boulanger2014,zhekova2014,filatov2014,guennic2015,minezawa2015,garniron2018}.

Here, we employ quantum Monte Carlo (QMC) to revisit the vertical excitation energies of cyanine
dyes of the simple form C$_n$H$_n$(NH$_2)_2^+$ with $n$ an odd number ranging from 1 to 17,
combining the use of sophisticated
multi-determinant wave functions with recent developments for their efficient optimization in variational
Monte Carlo (VMC)~\cite{sorella2010,neuscamman2012,filippi2016,assaraf2017}. In particular, we build
on our successful treatment at chemical accuracy of the excitation energies and optimal
excited-state structures of small, prototypical molecules~\cite{dash2019,cuzzocrea2020,dash2021}, where
the determinantal components of the multiple states are generated in an automatic and balanced manner
with the configuration interaction using a perturbative selection made iteratively (CIPSI) approach~\cite{huron1973}.
Studying the bright excitation of cyanine dyes enables us to demonstrate the 
accuracy of our protocol for the shorter chains, where high-level coupled cluster (CC) offers a 
good compromise in terms of accuracy versus computational cost. Importantly, it also establishes the
applicability of QMC to larger sizes where the use of other high-level approaches is more challenging.
Finally, we identify the key descriptors of orbital correlations for these systems and elucidate why earlier QMC studies
with limited active space wave functions lacked the expected accuracy~\cite{send2011}.


\section{Methods}
\label{sec:methods}

We employ QMC wave functions of the so-called Jastrow-Slater form, namely,
\begin{eqnarray}
\Psi = \mathcal{J} \sum_{i=1}^{N_{\rm det}} c_i D_i\,,
\end{eqnarray}
where $\mathcal{J}$ is the Jastrow correlation factor and $D_i$ are determinants of single-particle
orbitals. The Jastrow factor explicitly depends on the inter-particle coordinates and includes here
electron-electron and electron-nucleus correlation terms~\cite{Jastrow}.


To generate the determinantal components for the two states, we employ the
CIPSI approach which, starting from a given reference space, builds expansions by
iteratively selecting determinants based on their second-order perturbation (PT2) energy contribution
obtained via the Epstein-Nesbet partitioning of the Hamiltonian~\cite{epstein1926,nesbet1955},
\begin{eqnarray}
\delta E_{\alpha}^{(2)} = \frac{|\langle\alpha|\ham|\Psi^{\rm CIPSI}\rangle|^2}{\langle\Psi^{\rm CIPSI}|\ham|\Psi^{\rm CIPSI}\rangle - \langle\alpha|\ham|\alpha\rangle}\,,
\label{eq:en-pt2}
\end{eqnarray}
where $\Psi^{\rm CIPSI}$ is the current CIPSI wave function for the state under consideration
and $|\alpha\rangle$ denotes a  determinant outside the current CI space.  Since the ground and excited
states of the cyanines have different symmetries,
a state-specific approach can be used to perform the
selection for the two states separately, using different orbitals.

We are here interested in computing excitation energies and, therefore, wish to achieve a balanced
CIPSI description of the states of interest, which leads to converged excitation energies in QMC already
for relatively small expansions.
A measure
of the quality of a given CIPSI wave function is its PT2 energy contribution, which represents an
approximate estimate of the error of the expansion with respect to the full CI (FCI) limit. Therefore,
we can compute the excitation energies using expansions for the two (or more generally multiple) states with matched PT2 energy
and, therefore, ensure comparable quality.  We refer the reader to Ref.~\cite{dash2021} on how to
impose the ``iso-PT2'' criterion when treating multiple states of the same symmetry expanded
on a common set of determinants.

Alternatively, one can match the CI variance of the relevant states, which is defined as the variance
of the FCI Hamiltonian:
\begin{eqnarray}
\sigma^2_{\text{CI}} (\Psi^{\rm CIPSI}) &=&
\sum_{i \in \text{FCI}}
\langle \Psi^{\rm CIPSI} |\ham| i \rangle
\langle i|\ham|\Psi^{\rm CIPSI}\rangle   \nonumber \\
&&-\langle \Psi^{\rm CIPSI} |\ham| \Psi^{\rm CIPSI} \rangle ^ 2 \nonumber \\
&=& \sum_\alpha |\langle \Psi^{\rm CIPSI} |\ham| \alpha \rangle|^2
\,.
\label{eq:var_cipsi}
\end{eqnarray}
As the CIPSI wave function approaches the FCI limit, the CI variance goes to zero.  For various small
molecules~\cite{dash2019,cuzzocrea2020, dash2021}, we have found that matching the PT2 energy contributions
leads to expansions with also very similar variances. In general, this is not always the case and one
of the two criteria might be more suitable than the other for the computation of the CI excitation energies
of a particular system. 

While we discuss in detail below the impact of this choice on the QMC excitation energies, 
we stress already here that the convergence of the QMC results is established not based on their agreement
with available reference data but in an 
``internally consistent'' manner based on the similarity of the VMC and DMC excitation energies~\cite{dash2021}
and their convergence with respect to the number of determinants.

Finally, as an alternative to the CIPSI expansions, we test complete active space (CAS) expansions for
the determinantal components of our QMC wave functions. We start from separate CASSCF calculations
for the two states and consider minimal active spaces by correlating the $\pi$ electrons
in the $\pi$ orbitals constructed from the $2p_z$ orbitals. For the smaller cyanines with up to 7 heavy
atoms, CN3--CN7 (we label a cyanine as CN$m$ with $m$ the total number of heavy atoms), we also explore
the use of a larger active space with molecular orbitals constructed from the $2p_z$ and $3p_z$ atomic
orbitals.  Finally, in some cases, we also test the performance of a simple one-configuration ansatz,
namely, the Hartree-Fock (HF) and HOMO-LUMO (HL) configurations for the ground and the excited state,
respectively.

\section{Computational details}
\label{sec:comput}

Unless otherwise specified, we employ scalar-relativistic
energy-consistent HF pseudopotentials and the correlation-consistent Gaussian basis sets specifically
constructed for these pseudopotentials~\cite{burkatzki2007,BFD_H2013}.  For most of the calculations,
we use a double-$\zeta$ basis set minimally augmented with $s$ and $p$ diffuse functions on the heavy
atoms and denoted here as maug-cc-pVDZ.  Convergence tests are performed with the fully-augmented
aug-cc-pVTZ basis set.  The exponents of the diffuse functions are taken from the corresponding
all-electron Dunning's correlation-consistent basis sets~\cite{kendall1992}.

The HF and CASSCF computations are performed with the program GAMESS(US)~\cite{schmidt1993,gordon2005}.
When using the CASSCF wave functions in QMC, we truncate the CAS expansion for CN11 and CN13, using a
threshold on the CSF coefficients so that the configurations make up respectively about 0.9985 and
0.9765 of the weight of the total wave functions of the two states.
The CIPSI expansions are generated with Quantum Package~\cite{garniron2019}
and constructed to be eigenstates of ${\hat S}^{2}$~\cite{dash2019}. We perform the selection for the two states separately, starting from
CASSCF orbitals obtained with the larger active spaces for
the cyanine molecules up to CN7, and the minimal CAS from CN9 to CN15.
We use the HF orbitals for CN17 and CN19. As shown in Fig.~S1 
and Table~S5 for CN3 and Fig.~S6 for CN15, the use of different 
orbitals to generate the CIPSI expansions has no appreciable impact 
on the CI or QMC excitation energies.

The QMC calculations are carried out with the CHAMP code~\cite{Champ}.
The determinantal part of our QMC wave functions is
expressed in terms of spin-adapted configuration state functions
(CSF) to reduce the number of parameters during the VMC  optimization.  In
the wave function optimization, we sample a guiding wave function that differs from the current wave
function close to the nodes~\cite{attaccalite2008} to guarantee finite variances of the estimators of
the gradients with respect to the wave function parameters.
All wave
function parameters (Jastrow, CI, and orbital coefficients) are optimized in state-specific energy
minimization following the stochastic reconfiguration scheme~\cite{sorella2007,neuscamman2012}.
In the DMC calculations, we treat the pseudopotentials beyond the locality
approximation using the T-move algorithm~\cite{casula2006a} and employ an imaginary time-step of 0.05
a.u.\ which we have already tested for one of the cyanine chains and shown to yield excitation energies
converged to better than 0.01 eV~\cite{cuzzocrea2020}

We compute all energies on the ground-state geometries of CN3--CN11 determined with all-electron PBE0/cc-pVQZ 
in Ref.~\cite{boulanger2014} and obtain the geometries for CN13 to CN19 at the same level of theory 
with the Gaussian 09 program~\cite{gaussian09}.  We employ
the programs CFour v2.1~\cite{cfour} and Molcas~\cite{molcas8} for the approximate coupled cluster singles and
doubles (CC2) and singles, doubles, and triples model (CC3), and the CASPT2 calculations, respectively, using the all-electron
aug-cc-pVDZ basis set and the frozen-core approximation, unless otherwise specified.

\section{Results}
\label{sec:results}

We compute the lowest $\pi \rightarrow \pi^{*}$ vertical excitation energy of cyanine dyes of the form
C$_n$H$_n$(NH$_2)_2^+$ with $n$ ranging from 1 up to 17.
The structures of the CN3 and CN9 molecules
are shown in Figure~\ref{fig:cipsi_conv}. In all cases, the point group of the molecule is C$_{2v}$ with
the ground (GS) and excited (ES) states having A$_1$ and B$_1$ symmetry, respectively.

For CN3 up to CN15, we compare the QMC excitation energies with the all-electron CC3/aug-cc-pVDZ results.
The use of the CC3 method as reference for the bright excitation of these systems is supported by the
agreement of the CC3 excitation energies with the corresponding extrapolated FCI
(exFCI) estimates in a small basis of the smaller CN3 and CN5 to better than 0.05 eV~\cite{garniron2018}.
Employing the aug-cc-pVDZ basis set is sufficient given the
agreement with the corresponding aug-cc-pVTZ values (see
Table S1).  Importantly, the all-electron CC3/aug-cc-pVDZ excitation energies are very close
to the BFD CC3/aug-cc-pVDZ values, confirming that the use of pseudopotentials does
not introduce appreciable errors. The reference CC3/aug-cc-pVDZ values also agree with the
corresponding CC3 excitation energies computed with the BFD maug-cc-pVDZ basis set for all cyanines
except the smallest CN3 (see Table S1), where a fully-augmented double-$\zeta$ basis is needed
also in the BFD calculations.

For dyes larger than CN15, we are however not able to run
the CC3 calculations due to memory requirements~\cite{memCC} and the DMC excitation energy with our best
CIPSI wave function becomes then the reference for other calculations.

\subsection{Building the expansions}

To compute accurate QMC excitation energies for the cyanine dyes, one needs balanced Jastrow-Slater
wave functions to describe the ground and excited states. This is achieved in two stages, where the first
is the construction of CIPSI expansions with the iso-PT2 and/or iso-variance scheme, and the second is
a validation criterion that the resulting excitation energies in VMC and DMC 
are close to each other and converged with respect to the number of determinants.

In particular, we generate the ground- and excited-state expansions at the CIPSI level to have either
matched PT2 energy corrections or CI variances, which we use as measures of the ``distance'' of the
wave functions from the FCI limit. Imposing that the determinantal components satisfy either the iso-PT2
or iso-variance criterion was previously found to lead to QMC excitation energies which were converged
to the best reference values with a handful of determinants~\cite{dash2019,cuzzocrea2020}, even when
the error on the starting CI excitation energy was relatively large~\cite{dash2021}.

\begin{figure}[t]
\includegraphics[width=\columnwidth]{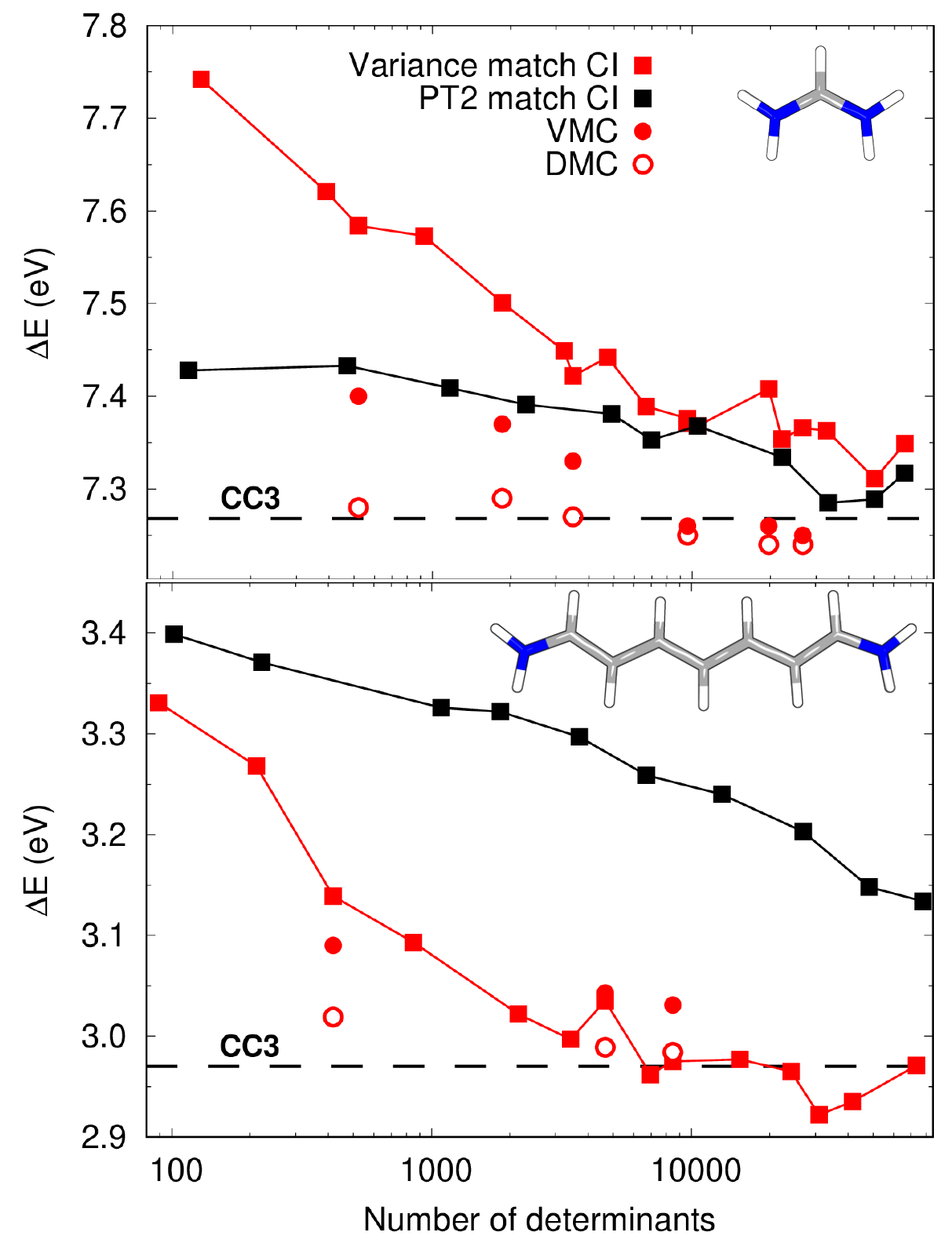}
\caption{CI vertical excitation energies of CN3 (top) and CN9 (bottom) versus the
total number of determinants, computed for ground- and excited-state CIPSI expansions having either matched
PT2 energy contributions or CI variances. The VMC and DMC excitation energies obtained using the iso-variance expansions
are also shown (the statistical error is smaller than the symbol size).
The BFD pseudopotentials and the maug-cc-pVDZ basis are used here also for the CC3 calculations.}
\label{fig:cipsi_conv}
\end{figure}

In Fig.~\ref{fig:cipsi_conv}, we illustrate the convergence of the CI excitation energies of CN3 and
CN9 versus the total number of determinants for expansions characterized by similar PT2 corrections
or CI variances.  For CN3, the iso-PT2 construction leads to a somewhat faster convergence of the
excitation energy for small expansions, but the two criteria become quickly equivalent beyond a few
1000 determinants.  The situation is reversed for CN9, where matching the PT2 correction yields a much
slower converging CI excitation energy, while the iso-variance criterion leads to a good agreement
with the CC3 value in the same basis set for little more than 1000 determinants. In fact, we find that
variance-matched expansions yield a faster converging CI excitation energy starting from CN7 and that,
surprisingly, fewer determinants are needed to obtain a good estimate for the larger system sizes
considered (see Fig.~S4). Consequently, the CI treatment of the smallest cyanine, CN3, appears to be the
most difficult as further elaborated in Sec.~\ref{orb-corr}.

Importantly, in Fig.~\ref{fig:cipsi_conv}, we also show that QMC largely corrects for possible
shortcomings of the starting CIPSI expansions, yielding excitation energies which display a rather
small dependence on the number of determinants, especially at the DMC level.  For CN3 and small
expansions, where the iso-variance criterion significantly overestimates the CI excitation energy, VMC
and DMC reduce the error at the CI level by about 0.2 and 0.3 eV, respectively. As the expansions
become larger, the difference between the VMC and the DMC values diminishes, falling well below chemical
accuracy (about 0.05 eV) for both CN3 and CN9.  The robustness of the QMC results is further corroborated
for CN7 in Table S6, where we show that, for comparable number of determinants, the use of PT2- and
variance-matched wave functions yields excitation energies which differ by about 0.2 eV at the CI level
but are very close in VMC and completely equivalent in DMC.

\subsection{Best QMC vertical excitations}

\begin{table*}[t]
\caption{Vertical excitation energies (eV) for the cyanine dyes computed with QMC and other highly-correlated methods.
The BFD pseudopotentials are used in QMC, while all other calculations are all-electron.
All energies are computed on PBE0/cc-pVQZ geometries. }
\begin{tabular}{llllllllllllllll}
\hline
Method & CN3$^a$ & CN5 & CN7 & CN9 & CN11 & CN13 & CN15 & CN17 & CN19 \\
\hline
VMC-CAS                    & 7.67(1)  & 5.13(1)  & 3.97(1) & 3.11(1) & 2.58(1) & 2.13(1) & --      & --      & -- \\
DMC-CAS                    & 7.49(1)  & 5.04(1)  & 3.83(1) & 3.04(1) & 2.55(1) & 2.15(1) & --      & --      & -- \\
VMC-CIPSI                  & 7.23(1)  & 4.83(1)  & 3.65(1) & 3.03(1) & 2.55(1) & 2.18(1) & 1.85(1) & 1.66(1) & 1.59(2) \\
DMC-CIPSI                  & 7.23(1)  & 4.86(1)  & 3.66(1) & 2.98(1) & 2.54(1) & 2.15(1) & 1.90(1) & 1.65(1) & 1.57(1)\\
CASPT2/aug-cc-pVDZ         & 6.94     & 4.64     & 3.56    & 2.91    & 2.45    & 2.11    & 1.85    & 1.65    & -- \\
CC2/aug-cc-oVDZ            & 7.29     & 4.97     & 3.80    & 3.10    & 2.64    & 2.30    & 2.04    & 1.84    & -- \\
CC3/aug-cc-pVDZ            & 7.20     & 4.85     & 3.67    & 2.97    & 2.50    & 2.16    & 1.91    & --      & -- \\
exFCI/aug-cc-pVDZ~\cite{garniron2018}
                           & 7.17(2)  & 4.89(2)  & --      &  --     &  --     & --      & --      & --      & -- \\
\hline
\multicolumn{10}{l}{$^a$ The QMC-CIPSI calculations for CN3 are performed with the aug-cc-pVTZ basis. } \\
\end{tabular}
\label{tab:all_cyanine}
\end{table*}

In Table~\ref{tab:all_cyanine}, we summarize the VMC and DMC excitation energies of all cyanine dyes
obtained with the largest CIPSI expansions of Table S6 and the iso-variance selection criterion.  We
also list the QMC and CASPT2 excitation energies computed with minimal CAS expansions, together
with our CC2 and CC3 results and the exFCI estimates from the literature~\cite{garniron2018}.  We refer the
reader to Table S6 for additional QMC calculations with different numbers of determinants in the
Jastrow-CIPSI wave functions.

\begin{figure}[b]
\includegraphics[width=\columnwidth]{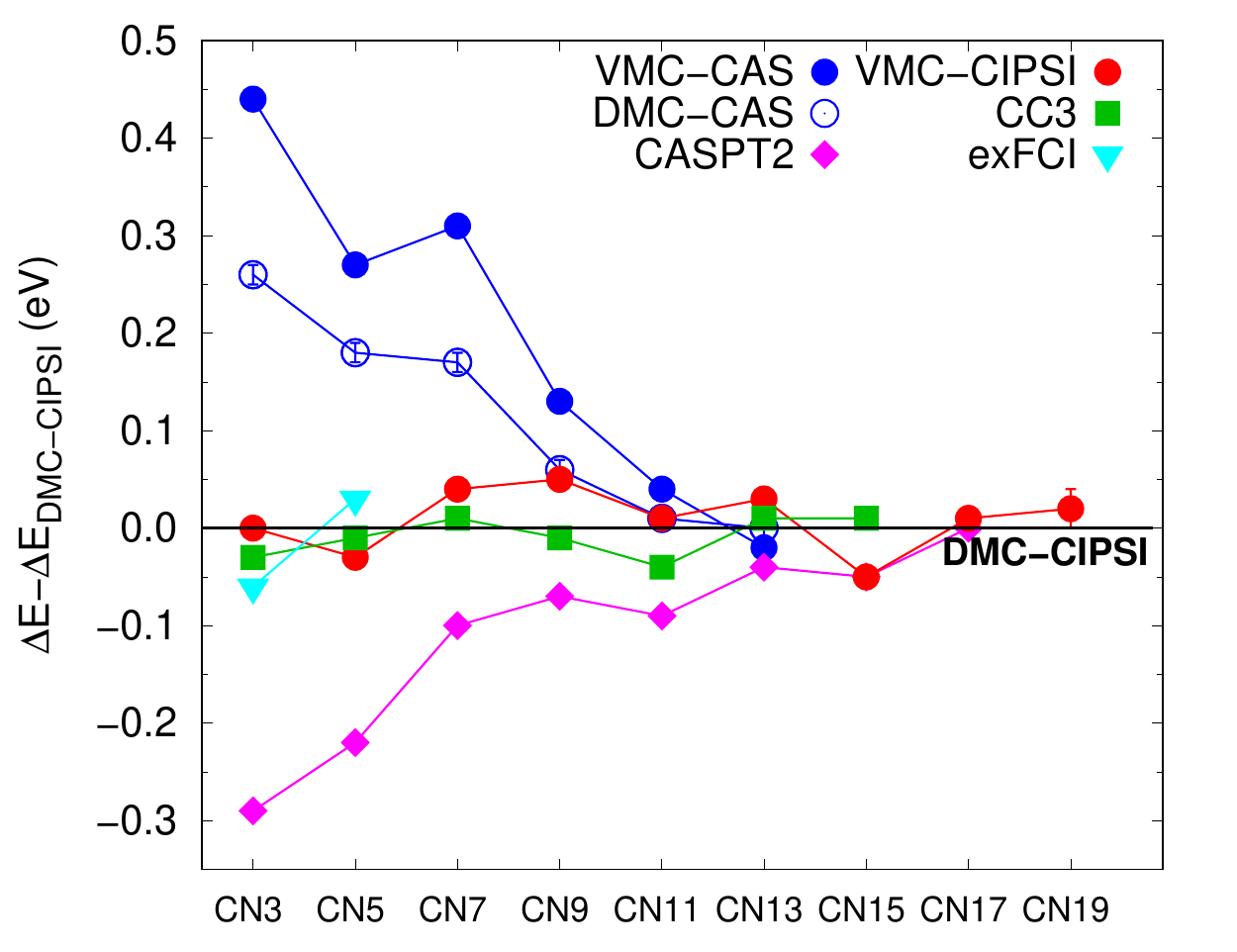}
\caption{Excitation energies (eV) at different levels of theory with respect to the DMC values
computed with the CIPSI wave functions (DMC-CIPSI line).}
\label{fig:err_cy}
\end{figure}

For the reported CIPSI expansions, the VMC and DMC excitations energies are very close and
also agree within chemical accuracy with the CC3 and exFCI values in all cases where these methods are
applicable.  This is in line with our previous findings that the agreement between VMC and DMC excitation
energy is a strong indication of the balanced quality of the corresponding wave functions~\cite{dash2021}.
Furthermore, we find that the QMC values for the larger dyes are in very good agreement with the
estimates given by the extrapolation of the CC3 results as a function of number of electrons
(see SI).
Since DMC can be employed in all cases, we plot all excitation energies in
Fig.~\ref{fig:err_cy} in terms of their distance to the DMC-CIPSI results, which we use as reference
values.

The CASPT2 and QMC-CAS energies computed with the minimal active spaces are instead very different from
the DMC-CIPSI results: CASPT2 always underestimates the excitation energies, whereas QMC-CAS tends to
overestimate them, similarly to what reported for CAS wave functions in Ref.~\cite{send2011}.
For CN3--CN7, we test the effect
of including more $\pi$ orbitals in the active space, which somewhat ameliorates the VMC excitation energies
but does not sufficiently affect the DMC values, which remain far from the DMC-CIPSI reference (see
Table S7).  Interestingly, we note that both the CASPT2 and QMC-CAS methods approach the best DMC results
as the size of the molecule increases, suggesting an easier treatment of the longer chains as already
found at the CI level and as further discussed below.

\subsection{Capturing orbital correlation}
\label{orb-corr}

\begin{figure}[!htb]
\includegraphics[width=\columnwidth]{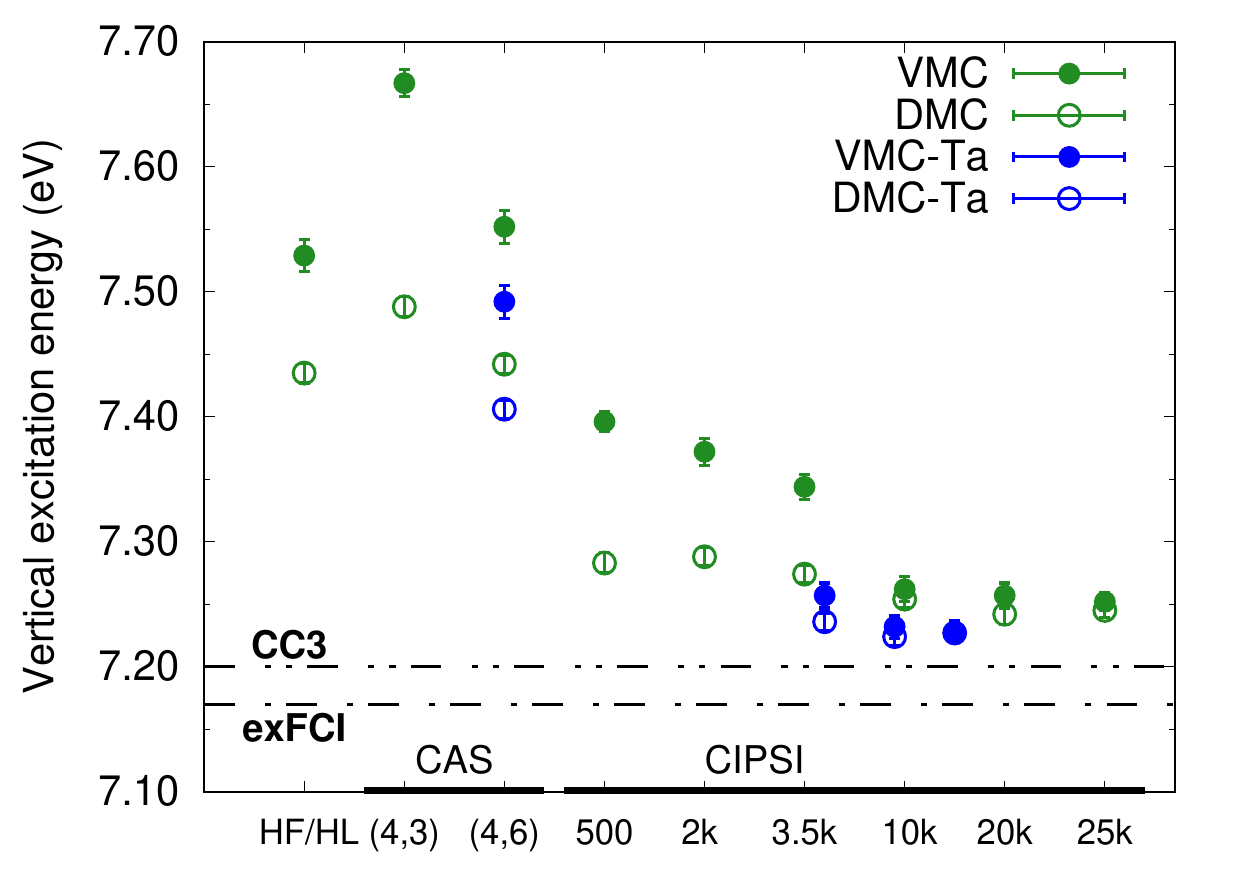}
\caption{VMC (full circle) and DMC (empty circle) vertical excitation energies of CN3 for different
wave functions.  The maug-cc-pVDZ (green) and aug-cc-pVTZ (Ta, blue) are used. }
\label{fig:cn3_ene}
\end{figure}

To understand the different performance of CAS and CIPSI expansions when used in QMC wave functions,
we focus here on CN3 and analyze in Fig.~\ref{fig:cn3_ene}
the VMC and DMC vertical excitation energies calculated using different determinantal
components in the trial wave functions.  As already mentioned, despite being the smallest cyanine dye,
CN3 appears to be the most challenging one:  the use of Jastrow-CAS wave functions leads to quite big
errors and the number of CIPSI determinants needed to converge the excitation energy is larger than for the longer dyes.

The simplest QMC calculations are performed with a one-configuration
(HF/HL) wave function and the maug-cc-pVDZ basis set. We then proceed to CAS determinantal components
and CIPSI expansions also employing the aug-cc-pVTZ basis set.
The VMC excitation energy computed with the minimal CAS wave functions is worse than the HF/HL value
since the active space comprises more determinants for the ground state but only the HL configuration
for the excited state. DMC ameliorates the result but using a larger CAS space on the $\pi$ orbitals only marginally
helps (see Table S7).  On the other hand, with the CIPSI selected determinants, we have a considerable
improvement on the excitation energy and, with the use of just few hundred determinants, the DMC error
reduces to less then 0.1~eV. Employing larger expansions with the maug-cc-pVDZ basis set, we finally
converge to VMC values which are consistent with the DMC ones and approximately 0.04 eV higher than
the reference. The use of the aug-cc-pVTZ basis set further reduces the excitation energy by about
0.02 eV. We note that, for the longer cyanine chains, the smaller maug-cc-pVDZ basis set is found to
be sufficient for the computation of this excitation energy~\cite{cuzzocrea2020}.

The superior performance of the use of a CIPSI with respect to the CAS expansions in QMC
indicates that some key descriptor of correlation is missing from the active space and
is not recovered through the addition of the Jastrow factor
and the subsequent full optimization in VMC, nor through a DMC calculation with the optimal
Jastrow-Slater wave function. In this work as in Ref.~\cite{send2011},
the active space is chosen to correlate the $\pi$ electrons in the $\pi$ orbitals. From the QMC-CIPSI
results, we can therefore infer that, while the excitation of interest is predominantly of $\pi\to\pi^*$
character, other orbital correlations are important and
cannot be omitted in the QMC wave function of the shorter cyanines.

\begin{table}[!htb]
\caption[]{CI total energies (a.u.) and vertical excitation energies ($\Delta$E$_{\rm exc}$, eV) of
CN3 computed with the 6-31G basis set and different orbital sets.
The last column reports the error with respect to the FCI excitation energy for CN3 and CN5, and with
respect to the CC3 value for CN7.}
\begin{tabular}{lcc@{\hspace*{1em}}lc}
\hline
                       &        E(GS)  &        E(ES)  & $\Delta$E$_{\rm exc}$  & err \\
\hline
CN3 \\
         1 CSF         &   -149.39966  &   -149.07223  &   8.91 &	1.39(2) \\
         CAS-$\sigma$  &   -149.613(1) &   -149.307(1) &   8.32 &	0.80(2)\\
         CAS-$\pi$     &   -149.44486  &   -149.14840  &   8.07 &	0.55(2)\\
         CAS-$\pi$ +
         SD-$\sigma$   &  -149.7151(5) &   -149.4346(5)&   7.65 &	0.13(2)\\
         CC3           &   -149.74049  &   -149.46354  &   7.54 &	0.02(2)\\
         FCI           &   -149.741(1) &   -149.465(1) &   7.52(2) & -- \\
\hline
CN5 \\
         1 CSF         &   -226.27705  &   -226.04468  &   6.32 &	1.48(1) \\
         CAS-$\sigma$  &   -226.581(1) &   -226.373(1) &   5.66 &	0.82(1) \\
         CAS-$\pi$     &   -226.34204  &   -226.15212  &   5.17 &	0.33(1)\\
         CAS-$\pi$ +
         SD-$\sigma$   &   -226.745(2) &   -226.557(2) &   5.03 &	0.19(1) \\
         CC3           &   -226.80736  &   -226.62972  &   4.83 &	-0.01(1) \\
         FCI           &   -226.809(1) &   -226.631(1) &   4.84(1) & --  \\
\hline
CN7 \\
         1 CSF         &   -303.14723  &   -302.95611  &   5.20     & 1.64 \\
         CAS-$\sigma$  &   -303.580(4) &   -303.409(4) &   4.73(4)  & 1.17(4) \\
         CAS-$\pi$     &   -303.23260  &   -303.09606  &   3.72     & 0.16\\
         CC3           &   -303.86766  &   -303.73676  &   3.56     & --  \\
\hline
\end{tabular}
\label{tab:cipsi_cn3}
\end{table}

To better understand this, we present a CI study for CN3--CN7 with the small 6-31G basis set in
Table~\ref{tab:cipsi_cn3}.  We correlate only the valence electrons and use state-average natural
orbitals obtained with a preliminary calculation at the CIPSI level.
For each state, we compute the energy with only one CSF and, on top of this configuration, we perform a
CAS-CI calculation restricted to the $\sigma$ and the $\pi$ orbitals in a CAS-$\sigma$ and
CAS-$\pi$, respectively.
The reference FCI excitation energy in this basis and the associated confidence interval are 
computed following the scheme presented in Ref.~\onlinecite{veril_2021} rather than 
extrapolating the variational energies of the individual states in the
limit of the PT2 energy correction going to zero. 
Indeed, the uncertainties of the 
extrapolated FCI energies of both states is larger than the uncertainty on the estimated 
excitation energy computed with this scheme.
Since the CC3 estimate for CN3 and CN5 is in excellent agreement with the FCI value, we use the CC3 excitation energy as reference for CN7.
We note that, because of the use of the simple 6-31G basis set, the FCI and CC3 excitation
energies are much higher than the more accurate results presented above but this is not relevant for the
present discussion.


For the CN3 molecule, the excitation energy obtained with a single CSF for each state is 8.91~eV,
namely, higher by 1.4~eV than the FCI result.  The CAS-$\pi$ calculation corrects only 60\% of the
error, indicating that the $\sigma$ orbitals also play an important role in the stabilization of the
excited state.  Similarly, the excitation energy obtained with the CAS-$\sigma$ improves the excitation
energy with respect to the single CSF by recovering about 31\% of the error. These results indicate
the importance of both $\sigma$ and $\pi$ orbitals in the calculation of the excitation energy of CN3.

Therefore, to partially account for both $\pi$ and $\sigma$ correlations,
we perform a multi-reference CI calculation, applying all possible single and double excitations to
the CAS-$\pi$ determinants. Such a  CAS-$\pi$+SD-$\sigma$ calculation also enables the
relaxation of the CAS-$\pi$ CI coefficients in the presence of most of the $\sigma$ correlation.
The resulting excitation energy is now significantly improved but still 0.1~eV higher than the FCI reference,
confirming that a similar computational effort needs to be made for the $\pi$ and $\sigma$ orbitals.
This justifies the use of
CIPSI where the most important Slater determinants will be chosen to describe $\sigma$, $\pi$, and $\sigma-\pi$
correlation in a ``democratic'' way based on their contribution to the second-order perturbation energy.

For CN5, the situation is somewhat different. While the single CSF still overestimates the excitation
energy by 1.47~eV, the CAS-$\pi$ wave function behaves better than for CN3, recovering 80\% of the error.
Consequently, omitting the $\sigma$ orbitals in the active space results in an
excitation energy closer to the reference than in the CN3 case. Once the $\sigma$ orbitals are introduced, as
for CN3, we improve the excitation energy but still observe an overestimation of the CAS-$\pi$+SD-$\sigma$ result
by almost 0.2~eV, pointing to the importance of describing the $\sigma$ as well as the $\pi$ correlation.
The situation for CN7 is similar as for CN5, suggesting that the whole series behaves
like CN5 and that CN3 is an exception because of the particularly small length of the chain.

\section{Conclusion}
\label{sec:concl}

We have presented a QMC benchmark study of the lowest vertical excitation energies
of cyanine chains.
We constructed the determinantal components of the Jastrow-Slater wave functions
through an automatic selected-CI procedure and obtained a balanced description of
the relevant states by ensuring similar quality of the corresponding expansions,
for instance by matching their CI variances.  With compact expansions of only a
few thousand determinants, upon optimization of all parameters in our wave functions,
we obtained QMC excitation energies which improve on the starting CI values and,
for the shorter chain lengths where CC3 calculations are feasible,  agree with the
CC3 results to chemical accuracy.  We also applied our protocol to longer cyanines
and validated the accuracy of our estimates via the consistent closeness of the
determined VMC and DMC excitation energies.  Finally, we showed that key to a
successful description of this excitation over all chain lengths is to account for
$\pi$, $\sigma$, and $\sigma-\pi$ correlations, therefore going beyond
a CAS treatment based on $\pi$-orbitals only. In conclusion, we believe that
the present study further establishes QMC methods as accurate and robust tools for
the treatment of excited states of relatively large systems and parameter spaces.

\section*{Acknowledgment}

A.C.\ is supported by the ``Computational Science for Energy Research and Netherlands
eScience Center joint program''
(project CSER.JCER.022) of the Netherlands Organisation for Scientific Research (NWO).
The calculations were carried out on the Dutch national supercomputer Cartesius
with the support of SURF Cooperative.
The work is also partially supported by
the European Center of Excellence in Exascale Computing TREX - Targeting Real
Chemical Accuracy at the Exascale, funded from the European Union's Horizon
2020 Research and Innovation program (grant no.\ 952165).

\section*{Content of SI}
Dependence of the excitation energies on basis set and pseudopotential;
detailed information on all CIPSI and QMC calculations for CN3;
CIPSI energies and excitation energies obtained by matching either the PT2
energy or the CI variance for all dyes; impact of using HF orbitals in
generating CIPSI expansions; QMC-CIPSI and QMC-CAS energies and excitation energies;
PBE0/cc-pVQZ geometries of CN13-CN19.

\bibliography{paper_cyanines_cipsi}

\begin{thebibliography}{38}%
\makeatletter
\providecommand \@ifxundefined [1]{%
 \@ifx{#1\undefined}
}%
\providecommand \@ifnum [1]{%
 \ifnum #1\expandafter \@firstoftwo
 \else \expandafter \@secondoftwo
 \fi
}%
\providecommand \@ifx [1]{%
 \ifx #1\expandafter \@firstoftwo
 \else \expandafter \@secondoftwo
 \fi
}%
\providecommand \natexlab [1]{#1}%
\providecommand \enquote  [1]{``#1''}%
\providecommand \bibnamefont  [1]{#1}%
\providecommand \bibfnamefont [1]{#1}%
\providecommand \citenamefont [1]{#1}%
\providecommand \href@noop [0]{\@secondoftwo}%
\providecommand \href [0]{\begingroup \@sanitize@url \@href}%
\providecommand \@href[1]{\@@startlink{#1}\@@href}%
\providecommand \@@href[1]{\endgroup#1\@@endlink}%
\providecommand \@sanitize@url [0]{\catcode `\\12\catcode `\$12\catcode
  `\&12\catcode `\#12\catcode `\^12\catcode `\_12\catcode `\%12\relax}%
\providecommand \@@startlink[1]{}%
\providecommand \@@endlink[0]{}%
\providecommand \url  [0]{\begingroup\@sanitize@url \@url }%
\providecommand \@url [1]{\endgroup\@href {#1}{\urlprefix }}%
\providecommand \urlprefix  [0]{URL }%
\providecommand \Eprint [0]{\href }%
\providecommand \doibase [0]{http://dx.doi.org/}%
\providecommand \selectlanguage [0]{\@gobble}%
\providecommand \bibinfo  [0]{\@secondoftwo}%
\providecommand \bibfield  [0]{\@secondoftwo}%
\providecommand \translation [1]{[#1]}%
\providecommand \BibitemOpen [0]{}%
\providecommand \bibitemStop [0]{}%
\providecommand \bibitemNoStop [0]{.\EOS\space}%
\providecommand \EOS [0]{\spacefactor3000\relax}%
\providecommand \BibitemShut  [1]{\csname bibitem#1\endcsname}%
\let\auto@bib@innerbib\@empty
\bibitem [{\citenamefont {Levitus}\ and\ \citenamefont
  {Ranjit}(2011)}]{levitus2011}%
  \BibitemOpen
  \bibfield  {author} {\bibinfo {author} {\bibfnamefont {M.}~\bibnamefont
  {Levitus}}\ and\ \bibinfo {author} {\bibfnamefont {S.}~\bibnamefont
  {Ranjit}},\ }\href {\doibase 10.1017/S0033583510000247} {\bibfield  {journal}
  {\bibinfo  {journal} {Q. Rev. Biophys.}\ }\textbf {\bibinfo {volume} {44}},\
  \bibinfo {pages} {123} (\bibinfo {year} {2011})}\BibitemShut {NoStop}%
\bibitem [{\citenamefont {Shindy}(2017)}]{shindy2017}%
  \BibitemOpen
  \bibfield  {author} {\bibinfo {author} {\bibfnamefont {H.}~\bibnamefont
  {Shindy}},\ }\href {\doibase https://doi.org/10.1016/j.dyepig.2017.06.029}
  {\bibfield  {journal} {\bibinfo  {journal} {Dyes Pigm.}\ }\textbf {\bibinfo
  {volume} {145}},\ \bibinfo {pages} {505 } (\bibinfo {year}
  {2017})}\BibitemShut {NoStop}%
\bibitem [{\citenamefont {Oprea}\ \emph {et~al.}(2020)\citenamefont {Oprea},
  \citenamefont {Panait}, \citenamefont {Essam}, \citenamefont {Abd El-Aal},\
  and\ \citenamefont {G\^{i}r\textcommabelow{t}u}}]{oprea2020}%
  \BibitemOpen
  \bibfield  {author} {\bibinfo {author} {\bibfnamefont {C.~I.}\ \bibnamefont
  {Oprea}}, \bibinfo {author} {\bibfnamefont {P.}~\bibnamefont {Panait}},
  \bibinfo {author} {\bibfnamefont {Z.~M.}\ \bibnamefont {Essam}}, \bibinfo
  {author} {\bibfnamefont {R.~M.}\ \bibnamefont {Abd El-Aal}}, \ and\ \bibinfo
  {author} {\bibfnamefont {M.~A.}\ \bibnamefont {G\^{i}r\textcommabelow{t}u}},\
  }\href {\doibase 10.3390/nano10040662} {\bibfield  {journal} {\bibinfo
  {journal} {Nanomaterials}\ }\textbf {\bibinfo {volume} {10}},\ \bibinfo
  {pages} {662} (\bibinfo {year} {2020})}\BibitemShut {NoStop}%
\bibitem [{\citenamefont {Send}\ \emph {et~al.}(2011)\citenamefont {Send},
  \citenamefont {Valsson},\ and\ \citenamefont {Filippi}}]{send2011}%
  \BibitemOpen
  \bibfield  {author} {\bibinfo {author} {\bibfnamefont {R.}~\bibnamefont
  {Send}}, \bibinfo {author} {\bibfnamefont {O.}~\bibnamefont {Valsson}}, \
  and\ \bibinfo {author} {\bibfnamefont {C.}~\bibnamefont {Filippi}},\ }\href
  {\doibase dx.doi.org/10.1021/ct1006295} {\bibfield  {journal} {\bibinfo
  {journal} {J. Chem. Theory Comput.}\ }\textbf {\bibinfo {volume} {7}},\
  \bibinfo {pages} {444} (\bibinfo {year} {2011})}\BibitemShut {NoStop}%
\bibitem [{\citenamefont {Jacquemin}\ \emph {et~al.}(2012)\citenamefont
  {Jacquemin}, \citenamefont {Zhao}, \citenamefont {Valero}, \citenamefont
  {Adamo}, \citenamefont {Ciofini},\ and\ \citenamefont
  {Truhlar}}]{jacquemin2012}%
  \BibitemOpen
  \bibfield  {author} {\bibinfo {author} {\bibfnamefont {D.}~\bibnamefont
  {Jacquemin}}, \bibinfo {author} {\bibfnamefont {Z.}~\bibnamefont {Zhao}},
  \bibinfo {author} {\bibfnamefont {R.}~\bibnamefont {Valero}}, \bibinfo
  {author} {\bibfnamefont {C.}~\bibnamefont {Adamo}}, \bibinfo {author}
  {\bibfnamefont {I.}~\bibnamefont {Ciofini}}, \ and\ \bibinfo {author}
  {\bibfnamefont {D.}~\bibnamefont {Truhlar}},\ }\href {\doibase
  dx.doi.org/10.1021/ct200721d} {\bibfield  {journal} {\bibinfo  {journal} {J.
  Chem. Theory Comput.}\ }\textbf {\bibinfo {volume} {8}},\ \bibinfo {pages}
  {1255} (\bibinfo {year} {2012})}\BibitemShut {NoStop}%
\bibitem [{\citenamefont {Moore}\ and\ \citenamefont
  {Autschbach}(2013)}]{moore2013}%
  \BibitemOpen
  \bibfield  {author} {\bibinfo {author} {\bibfnamefont {B.}~\bibnamefont
  {Moore}}\ and\ \bibinfo {author} {\bibfnamefont {J.}~\bibnamefont
  {Autschbach}},\ }\href {\doibase 10.1021/ct400649r} {\bibfield  {journal}
  {\bibinfo  {journal} {J. Chem. Theory Comput.}\ }\textbf {\bibinfo {volume}
  {9}},\ \bibinfo {pages} {4991} (\bibinfo {year} {2013})}\BibitemShut
  {NoStop}%
\bibitem [{\citenamefont {Boulanger}\ \emph {et~al.}(2014)\citenamefont
  {Boulanger}, \citenamefont {Jacquemin}, \citenamefont {Duchemin},\ and\
  \citenamefont {Blase}}]{boulanger2014}%
  \BibitemOpen
  \bibfield  {author} {\bibinfo {author} {\bibfnamefont {P.}~\bibnamefont
  {Boulanger}}, \bibinfo {author} {\bibfnamefont {D.}~\bibnamefont
  {Jacquemin}}, \bibinfo {author} {\bibfnamefont {I.}~\bibnamefont {Duchemin}},
  \ and\ \bibinfo {author} {\bibfnamefont {X.}~\bibnamefont {Blase}},\ }\href
  {\doibase dx.doi.org/10.1021/ct401101u} {\bibfield  {journal} {\bibinfo
  {journal} {J. Chem. Theory Comput.}\ }\textbf {\bibinfo {volume} {10}},\
  \bibinfo {pages} {1212} (\bibinfo {year} {2014})}\BibitemShut {NoStop}%
\bibitem [{\citenamefont {Zhekova}\ \emph {et~al.}(2014)\citenamefont
  {Zhekova}, \citenamefont {Krykunov}, \citenamefont {Autschbach},\ and\
  \citenamefont {Ziegler}}]{zhekova2014}%
  \BibitemOpen
  \bibfield  {author} {\bibinfo {author} {\bibfnamefont {H.}~\bibnamefont
  {Zhekova}}, \bibinfo {author} {\bibfnamefont {M.}~\bibnamefont {Krykunov}},
  \bibinfo {author} {\bibfnamefont {J.}~\bibnamefont {Autschbach}}, \ and\
  \bibinfo {author} {\bibfnamefont {T.}~\bibnamefont {Ziegler}},\ }\href
  {\doibase 10.1021/ct500292c} {\bibfield  {journal} {\bibinfo  {journal} {J.
  Chem. Theory Comput.}\ }\textbf {\bibinfo {volume} {10}},\ \bibinfo {pages}
  {3299} (\bibinfo {year} {2014})}\BibitemShut {NoStop}%
\bibitem [{\citenamefont {Filatov}\ and\ \citenamefont
  {Huix-Rotllant}(2014)}]{filatov2014}%
  \BibitemOpen
  \bibfield  {author} {\bibinfo {author} {\bibfnamefont {M.}~\bibnamefont
  {Filatov}}\ and\ \bibinfo {author} {\bibfnamefont {M.}~\bibnamefont
  {Huix-Rotllant}},\ }\href {\doibase 10.1063/1.4887087} {\bibfield  {journal}
  {\bibinfo  {journal} {J. Chem. Phys.}\ }\textbf {\bibinfo {volume} {141}},\
  \bibinfo {pages} {024112} (\bibinfo {year} {2014})}\BibitemShut {NoStop}%
\bibitem [{\citenamefont {Boris}\ and\ \citenamefont
  {Denis}(2015)}]{guennic2015}%
  \BibitemOpen
  \bibfield  {author} {\bibinfo {author} {\bibfnamefont {L.~G.}\ \bibnamefont
  {Boris}}\ and\ \bibinfo {author} {\bibfnamefont {J.}~\bibnamefont {Denis}},\
  }\href {\doibase https://doi.org/10.1021/ar500447q} {\bibfield  {journal}
  {\bibinfo  {journal} {Acc. Chem. Res.}\ }\textbf {\bibinfo {volume} {48}},\
  \bibinfo {pages} {530} (\bibinfo {year} {2015})}\BibitemShut {NoStop}%
\bibitem [{\citenamefont {Minezawa}(2015)}]{minezawa2015}%
  \BibitemOpen
  \bibfield  {author} {\bibinfo {author} {\bibfnamefont {N.}~\bibnamefont
  {Minezawa}},\ }\href {\doibase https://doi.org/10.1016/j.cplett.2015.01.033}
  {\bibfield  {journal} {\bibinfo  {journal} {Chem. Phys. Lett.}\ }\textbf
  {\bibinfo {volume} {622}},\ \bibinfo {pages} {115 } (\bibinfo {year}
  {2015})}\BibitemShut {NoStop}%
\bibitem [{\citenamefont {Yann}\ \emph {et~al.}(2018)\citenamefont {Yann},
  \citenamefont {Anthony}, \citenamefont {Emmanuel}, \citenamefont {Michel},\
  and\ \citenamefont {Loos}}]{garniron2018}%
  \BibitemOpen
  \bibfield  {author} {\bibinfo {author} {\bibfnamefont {G.}~\bibnamefont
  {Yann}}, \bibinfo {author} {\bibfnamefont {S.}~\bibnamefont {Anthony}},
  \bibinfo {author} {\bibfnamefont {G.}~\bibnamefont {Emmanuel}}, \bibinfo
  {author} {\bibfnamefont {C.}~\bibnamefont {Michel}}, \ and\ \bibinfo {author}
  {\bibfnamefont {P.-F.}\ \bibnamefont {Loos}},\ }\href {\doibase
  https://doi.org/10.1063/1.5044503} {\bibfield  {journal} {\bibinfo  {journal}
  {J. Chem. Phys.}\ }\textbf {\bibinfo {volume} {149}},\ \bibinfo {pages}
  {064103} (\bibinfo {year} {2018})}\BibitemShut {NoStop}%
\bibitem [{\citenamefont {Sorella}\ and\ \citenamefont
  {Capriotti}(2010)}]{sorella2010}%
  \BibitemOpen
  \bibfield  {author} {\bibinfo {author} {\bibfnamefont {S.}~\bibnamefont
  {Sorella}}\ and\ \bibinfo {author} {\bibfnamefont {L.}~\bibnamefont
  {Capriotti}},\ }\href {\doibase 10.1063/1.3516208} {\bibfield  {journal}
  {\bibinfo  {journal} {J. Chem. Phys.}\ }\textbf {\bibinfo {volume} {133}},\
  \bibinfo {pages} {234111} (\bibinfo {year} {2010})}\BibitemShut {NoStop}%
\bibitem [{\citenamefont {Neuscamman}\ \emph {et~al.}(2012)\citenamefont
  {Neuscamman}, \citenamefont {Umrigar},\ and\ \citenamefont
  {Chan}}]{neuscamman2012}%
  \BibitemOpen
  \bibfield  {author} {\bibinfo {author} {\bibfnamefont {E.}~\bibnamefont
  {Neuscamman}}, \bibinfo {author} {\bibfnamefont {C.~J.}\ \bibnamefont
  {Umrigar}}, \ and\ \bibinfo {author} {\bibfnamefont {G.~K.-L.}\ \bibnamefont
  {Chan}},\ }\href@noop {} {\bibfield  {journal} {\bibinfo  {journal} {Phys.
  Rev. B}\ }\textbf {\bibinfo {volume} {85}},\ \bibinfo {pages} {045103}
  (\bibinfo {year} {2012})}\BibitemShut {NoStop}%
\bibitem [{\citenamefont {Filippi}\ \emph {et~al.}(2016)\citenamefont
  {Filippi}, \citenamefont {Assaraf},\ and\ \citenamefont
  {Moroni}}]{filippi2016}%
  \BibitemOpen
  \bibfield  {author} {\bibinfo {author} {\bibfnamefont {C.}~\bibnamefont
  {Filippi}}, \bibinfo {author} {\bibfnamefont {R.}~\bibnamefont {Assaraf}}, \
  and\ \bibinfo {author} {\bibfnamefont {S.}~\bibnamefont {Moroni}},\
  }\href@noop {} {\bibfield  {journal} {\bibinfo  {journal} {J. Chem. Phys.}\
  }\textbf {\bibinfo {volume} {144}},\ \bibinfo {pages} {194105} (\bibinfo
  {year} {2016})}\BibitemShut {NoStop}%
\bibitem [{\citenamefont {Assaraf}\ \emph {et~al.}(2017)\citenamefont
  {Assaraf}, \citenamefont {Moroni},\ and\ \citenamefont
  {Filippi}}]{assaraf2017}%
  \BibitemOpen
  \bibfield  {author} {\bibinfo {author} {\bibfnamefont {R.}~\bibnamefont
  {Assaraf}}, \bibinfo {author} {\bibfnamefont {S.}~\bibnamefont {Moroni}}, \
  and\ \bibinfo {author} {\bibfnamefont {C.}~\bibnamefont {Filippi}},\
  }\href@noop {} {\bibfield  {journal} {\bibinfo  {journal} {J. Chem. Theory
  Comput.}\ }\textbf {\bibinfo {volume} {13}},\ \bibinfo {pages} {5273}
  (\bibinfo {year} {2017})}\BibitemShut {NoStop}%
\bibitem [{\citenamefont {Dash}\ \emph {et~al.}(2019)\citenamefont {Dash},
  \citenamefont {Feldt}, \citenamefont {Moroni}, \citenamefont {Scemama},\ and\
  \citenamefont {Filippi}}]{dash2019}%
  \BibitemOpen
  \bibfield  {author} {\bibinfo {author} {\bibfnamefont {M.}~\bibnamefont
  {Dash}}, \bibinfo {author} {\bibfnamefont {J.}~\bibnamefont {Feldt}},
  \bibinfo {author} {\bibfnamefont {S.}~\bibnamefont {Moroni}}, \bibinfo
  {author} {\bibfnamefont {A.}~\bibnamefont {Scemama}}, \ and\ \bibinfo
  {author} {\bibfnamefont {C.}~\bibnamefont {Filippi}},\ }\href {\doibase
  10.1021/acs.jctc.9b00476} {\bibfield  {journal} {\bibinfo  {journal} {J.
  Chem. Theory Comput.}\ }\textbf {\bibinfo {volume} {15}},\ \bibinfo {pages}
  {4896} (\bibinfo {year} {2019})}\BibitemShut {NoStop}%
\bibitem [{\citenamefont {Cuzzocrea}\ \emph {et~al.}(2020)\citenamefont
  {Cuzzocrea}, \citenamefont {Scemama}, \citenamefont {Briels}, \citenamefont
  {Moroni},\ and\ \citenamefont {Filippi}}]{cuzzocrea2020}%
  \BibitemOpen
  \bibfield  {author} {\bibinfo {author} {\bibfnamefont {A.}~\bibnamefont
  {Cuzzocrea}}, \bibinfo {author} {\bibfnamefont {A.}~\bibnamefont {Scemama}},
  \bibinfo {author} {\bibfnamefont {W.~J.}\ \bibnamefont {Briels}}, \bibinfo
  {author} {\bibfnamefont {S.}~\bibnamefont {Moroni}}, \ and\ \bibinfo {author}
  {\bibfnamefont {C.}~\bibnamefont {Filippi}},\ }\href {\doibase
  10.1021/acs.jctc.0c00147} {\bibfield  {journal} {\bibinfo  {journal} {J.
  Chem. Theory Comput.}\ }\textbf {\bibinfo {volume} {16}},\ \bibinfo {pages}
  {4203} (\bibinfo {year} {2020})}\BibitemShut {NoStop}%
\bibitem [{\citenamefont {Dash}\ \emph {et~al.}(2021)\citenamefont {Dash},
  \citenamefont {Moroni}, \citenamefont {Filippi},\ and\ \citenamefont
  {Scemama}}]{dash2021}%
  \BibitemOpen
  \bibfield  {author} {\bibinfo {author} {\bibfnamefont {M.}~\bibnamefont
  {Dash}}, \bibinfo {author} {\bibfnamefont {S.}~\bibnamefont {Moroni}},
  \bibinfo {author} {\bibfnamefont {C.}~\bibnamefont {Filippi}}, \ and\
  \bibinfo {author} {\bibfnamefont {A.}~\bibnamefont {Scemama}},\ }\href
  {\doibase 10.1021/acs.jctc.1c00212} {\bibfield  {journal} {\bibinfo
  {journal} {J. Chem. Theory Comput.}\ }\textbf {\bibinfo {volume} {17}},\
  \bibinfo {pages} {3426} (\bibinfo {year} {2021})}\BibitemShut {NoStop}%
\bibitem [{\citenamefont {Huron}\ \emph {et~al.}(1973)\citenamefont {Huron},
  \citenamefont {Malrieu},\ and\ \citenamefont {Rancurel}}]{huron1973}%
  \BibitemOpen
  \bibfield  {author} {\bibinfo {author} {\bibfnamefont {B.}~\bibnamefont
  {Huron}}, \bibinfo {author} {\bibfnamefont {J.}~\bibnamefont {Malrieu}}, \
  and\ \bibinfo {author} {\bibfnamefont {P.}~\bibnamefont {Rancurel}},\
  }\href@noop {} {\bibfield  {journal} {\bibinfo  {journal} {J. Chem. Phys.}\
  }\textbf {\bibinfo {volume} {58}},\ \bibinfo {pages} {5745} (\bibinfo {year}
  {1973})}\BibitemShut {NoStop}%
\bibitem [{Jas()}]{Jastrow}%
  \BibitemOpen
  \href@noop {} {}\bibinfo {note} {As Jastrow factor, we use the exponential of
  the sum of two fifth-order polynomials of the electron-nuclear and the
  electron-electron distances, respectively, and rescale the inter-particle
  distances as $R=(1-\exp(-\kappa r))/\kappa$ with $\kappa$ set to 0.6 a.u. We
  employ different electron-nucleus Jastrow factors to describe the correlation
  of an elecron with C, N, and H. The total number of free parameters to be
  optimized in the Jastrow factor is 17 for the systems considered
  here.}\BibitemShut {Stop}%
\bibitem [{\citenamefont {Epstein}(1926)}]{epstein1926}%
  \BibitemOpen
  \bibfield  {author} {\bibinfo {author} {\bibfnamefont {P.~S.}\ \bibnamefont
  {Epstein}},\ }\href@noop {} {\bibfield  {journal} {\bibinfo  {journal} {Phys.
  Rev.}\ }\textbf {\bibinfo {volume} {28}},\ \bibinfo {pages} {695} (\bibinfo
  {year} {1926})}\BibitemShut {NoStop}%
\bibitem [{\citenamefont {Nesbet}(1955)}]{nesbet1955}%
  \BibitemOpen
  \bibfield  {author} {\bibinfo {author} {\bibfnamefont {R.~K.}\ \bibnamefont
  {Nesbet}},\ }in\ \href@noop {} {\emph {\bibinfo {booktitle} {Proc. R. Soc.
  Lond. A}}},\ Vol.\ \bibinfo {volume} {230}\ (\bibinfo {organization} {The
  Royal Society},\ \bibinfo {year} {1955})\ pp.\ \bibinfo {pages}
  {312--321}\BibitemShut {NoStop}%
\bibitem [{\citenamefont {Burkatzki}\ \emph {et~al.}(2007)\citenamefont
  {Burkatzki}, \citenamefont {Filippi},\ and\ \citenamefont
  {Dolg}}]{burkatzki2007}%
  \BibitemOpen
  \bibfield  {author} {\bibinfo {author} {\bibfnamefont {M.}~\bibnamefont
  {Burkatzki}}, \bibinfo {author} {\bibfnamefont {C.}~\bibnamefont {Filippi}},
  \ and\ \bibinfo {author} {\bibfnamefont {M.}~\bibnamefont {Dolg}},\
  }\href@noop {} {\bibfield  {journal} {\bibinfo  {journal} {J. Chem. Phys.}\
  }\textbf {\bibinfo {volume} {126}},\ \bibinfo {pages} {234105} (\bibinfo
  {year} {2007})}\BibitemShut {NoStop}%
\bibitem [{BFD()}]{BFD_H2013}%
  \BibitemOpen
  \href@noop {} {}\bibinfo {note} {For the hydrogen atom, we use a more
  accurate BFD pseudopotential and basis set. Dolg, M.; Filippi, C., private
  communication}\BibitemShut {NoStop}%
\bibitem [{\citenamefont {Kendall}\ \emph {et~al.}(1992)\citenamefont
  {Kendall}, \citenamefont {Dunning~Jr},\ and\ \citenamefont
  {Harrison}}]{kendall1992}%
  \BibitemOpen
  \bibfield  {author} {\bibinfo {author} {\bibfnamefont {R.~A.}\ \bibnamefont
  {Kendall}}, \bibinfo {author} {\bibfnamefont {T.~H.}\ \bibnamefont
  {Dunning~Jr}}, \ and\ \bibinfo {author} {\bibfnamefont {R.~J.}\ \bibnamefont
  {Harrison}},\ }\href@noop {} {\bibfield  {journal} {\bibinfo  {journal} {J.
  Chem. Phys.}\ }\textbf {\bibinfo {volume} {96}},\ \bibinfo {pages} {6796}
  (\bibinfo {year} {1992})}\BibitemShut {NoStop}%
\bibitem [{\citenamefont {Schmidt}\ \emph {et~al.}(1993)\citenamefont
  {Schmidt}, \citenamefont {Baldridge}, \citenamefont {Boatz}, \citenamefont
  {Elbert}, \citenamefont {Gordon}, \citenamefont {Jensen}, \citenamefont
  {Koseki}, \citenamefont {Matsunaga}, \citenamefont {Nguyen}, \citenamefont
  {Su},\ and\ \citenamefont {{others}}}]{schmidt1993}%
  \BibitemOpen
  \bibfield  {author} {\bibinfo {author} {\bibfnamefont {M.~W.}\ \bibnamefont
  {Schmidt}}, \bibinfo {author} {\bibfnamefont {K.~K.}\ \bibnamefont
  {Baldridge}}, \bibinfo {author} {\bibfnamefont {J.~A.}\ \bibnamefont
  {Boatz}}, \bibinfo {author} {\bibfnamefont {S.~T.}\ \bibnamefont {Elbert}},
  \bibinfo {author} {\bibfnamefont {M.~S.}\ \bibnamefont {Gordon}}, \bibinfo
  {author} {\bibfnamefont {J.~H.}\ \bibnamefont {Jensen}}, \bibinfo {author}
  {\bibfnamefont {S.}~\bibnamefont {Koseki}}, \bibinfo {author} {\bibfnamefont
  {N.}~\bibnamefont {Matsunaga}}, \bibinfo {author} {\bibfnamefont {K.~A.}\
  \bibnamefont {Nguyen}}, \bibinfo {author} {\bibfnamefont {S.}~\bibnamefont
  {Su}}, \ and\ \bibinfo {author} {\bibnamefont {{others}}},\ }\href@noop {}
  {\bibfield  {journal} {\bibinfo  {journal} {J. Comput. Chem.}\ }\textbf
  {\bibinfo {volume} {14}},\ \bibinfo {pages} {1347} (\bibinfo {year}
  {1993})}\BibitemShut {NoStop}%
\bibitem [{\citenamefont {Gordon}\ and\ \citenamefont
  {Schmidt}(2005)}]{gordon2005}%
  \BibitemOpen
  \bibfield  {author} {\bibinfo {author} {\bibfnamefont {M.~S.}\ \bibnamefont
  {Gordon}}\ and\ \bibinfo {author} {\bibfnamefont {M.~W.}\ \bibnamefont
  {Schmidt}},\ }in\ \href@noop {} {\emph {\bibinfo {booktitle} {Theory and
  applications of computational chemistry}}}\ (\bibinfo  {publisher}
  {Elsevier},\ \bibinfo {year} {2005})\ pp.\ \bibinfo {pages}
  {1167--1189}\BibitemShut {NoStop}%
\bibitem [{\citenamefont {Garniron}\ \emph {et~al.}(2019)\citenamefont
  {Garniron}, \citenamefont {Applencourt}, \citenamefont {Gasperich},
  \citenamefont {Benali}, \citenamefont {Fert\'e}, \citenamefont {Paquier},
  \citenamefont {Pradines}, \citenamefont {Assaraf}, \citenamefont {Reinhardt},
  \citenamefont {Toulouse}, \citenamefont {Barbaresco}, \citenamefont {Renon},
  \citenamefont {David}, \citenamefont {Malrieu}, \citenamefont {V\'eril},
  \citenamefont {Caffarel}, \citenamefont {Loos}, \citenamefont {Giner},\ and\
  \citenamefont {Scemama}}]{garniron2019}%
  \BibitemOpen
  \bibfield  {author} {\bibinfo {author} {\bibfnamefont {Y.}~\bibnamefont
  {Garniron}}, \bibinfo {author} {\bibfnamefont {T.}~\bibnamefont
  {Applencourt}}, \bibinfo {author} {\bibfnamefont {K.}~\bibnamefont
  {Gasperich}}, \bibinfo {author} {\bibfnamefont {A.}~\bibnamefont {Benali}},
  \bibinfo {author} {\bibfnamefont {A.}~\bibnamefont {Fert\'e}}, \bibinfo
  {author} {\bibfnamefont {J.}~\bibnamefont {Paquier}}, \bibinfo {author}
  {\bibfnamefont {B.}~\bibnamefont {Pradines}}, \bibinfo {author}
  {\bibfnamefont {R.}~\bibnamefont {Assaraf}}, \bibinfo {author} {\bibfnamefont
  {P.}~\bibnamefont {Reinhardt}}, \bibinfo {author} {\bibfnamefont
  {J.}~\bibnamefont {Toulouse}}, \bibinfo {author} {\bibfnamefont
  {P.}~\bibnamefont {Barbaresco}}, \bibinfo {author} {\bibfnamefont
  {N.}~\bibnamefont {Renon}}, \bibinfo {author} {\bibfnamefont
  {G.}~\bibnamefont {David}}, \bibinfo {author} {\bibfnamefont {J.-P.}\
  \bibnamefont {Malrieu}}, \bibinfo {author} {\bibfnamefont {M.}~\bibnamefont
  {V\'eril}}, \bibinfo {author} {\bibfnamefont {M.}~\bibnamefont {Caffarel}},
  \bibinfo {author} {\bibfnamefont {P.-F.}\ \bibnamefont {Loos}}, \bibinfo
  {author} {\bibfnamefont {E.}~\bibnamefont {Giner}}, \ and\ \bibinfo {author}
  {\bibfnamefont {A.}~\bibnamefont {Scemama}},\ }\href {\doibase
  10.1021/acs.jctc.9b00176} {\bibfield  {journal} {\bibinfo  {journal} {J.
  Chem. Theory Comput.}\ }\textbf {\bibinfo {volume} {15}},\ \bibinfo {pages}
  {3591} (\bibinfo {year} {2019})}\BibitemShut {NoStop}%
\bibitem [{Cha()}]{Champ}%
  \BibitemOpen
  \href@noop {} {}\bibinfo {note} {CHAMP is a quantum Monte Carlo program
  package written by C. J. Umrigar, C. Filippi, S. Moroni and
  collaborators}\BibitemShut {NoStop}%
\bibitem [{\citenamefont {Attaccalite}\ and\ \citenamefont
  {Sorella}(2008)}]{attaccalite2008}%
  \BibitemOpen
  \bibfield  {author} {\bibinfo {author} {\bibfnamefont {C.}~\bibnamefont
  {Attaccalite}}\ and\ \bibinfo {author} {\bibfnamefont {S.}~\bibnamefont
  {Sorella}},\ }\href@noop {} {\bibfield  {journal} {\bibinfo  {journal} {Phys.
  Rev. Lett.}\ }\textbf {\bibinfo {volume} {100}},\ \bibinfo {pages} {114501}
  (\bibinfo {year} {2008})}\BibitemShut {NoStop}%
\bibitem [{\citenamefont {Sorella}\ \emph {et~al.}(2007)\citenamefont
  {Sorella}, \citenamefont {Casula},\ and\ \citenamefont
  {Rocca}}]{sorella2007}%
  \BibitemOpen
  \bibfield  {author} {\bibinfo {author} {\bibfnamefont {S.}~\bibnamefont
  {Sorella}}, \bibinfo {author} {\bibfnamefont {M.}~\bibnamefont {Casula}}, \
  and\ \bibinfo {author} {\bibfnamefont {D.}~\bibnamefont {Rocca}},\
  }\href@noop {} {\bibfield  {journal} {\bibinfo  {journal} {J. Chem. Phys.}\
  }\textbf {\bibinfo {volume} {127}},\ \bibinfo {pages} {014105} (\bibinfo
  {year} {2007})}\BibitemShut {NoStop}%
\bibitem [{\citenamefont {Casula}(2006)}]{casula2006a}%
  \BibitemOpen
  \bibfield  {author} {\bibinfo {author} {\bibfnamefont {M.}~\bibnamefont
  {Casula}},\ }\href {\doibase 10.1103/PhysRevB.74.161102} {\bibfield
  {journal} {\bibinfo  {journal} {Phys. Rev. B}\ }\textbf {\bibinfo {volume}
  {74}},\ \bibinfo {eid} {161102} (\bibinfo {year} {2006})}\BibitemShut
  {NoStop}%
\bibitem [{\citenamefont {Frisch}\ \emph {et~al.}(2016)\citenamefont {Frisch},
  \citenamefont {Trucks}, \citenamefont {Schlegel}, \citenamefont {Scuseria},
  \citenamefont {Robb}, \citenamefont {Cheeseman}, \citenamefont {Scalmani},
  \citenamefont {Barone}, \citenamefont {Petersson}, \citenamefont {Nakatsuji},
  \citenamefont {Li}, \citenamefont {Caricato}, \citenamefont {Marenich},
  \citenamefont {Bloino}, \citenamefont {Janesko}, \citenamefont {Gomperts},
  \citenamefont {Mennucci}, \citenamefont {Hratchian}, \citenamefont {Ortiz},
  \citenamefont {Izmaylov}, \citenamefont {Sonnenberg}, \citenamefont
  {Williams-Young}, \citenamefont {Ding}, \citenamefont {Lipparini},
  \citenamefont {Egidi}, \citenamefont {Goings}, \citenamefont {Peng},
  \citenamefont {Petrone}, \citenamefont {Henderson}, \citenamefont
  {Ranasinghe}, \citenamefont {Zakrzewski}, \citenamefont {Gao}, \citenamefont
  {Rega}, \citenamefont {Zheng}, \citenamefont {Liang}, \citenamefont {Hada},
  \citenamefont {Ehara}, \citenamefont {Toyota}, \citenamefont {Fukuda},
  \citenamefont {Hasegawa}, \citenamefont {Ishida}, \citenamefont {Nakajima},
  \citenamefont {Honda}, \citenamefont {Kitao}, \citenamefont {Nakai},
  \citenamefont {Vreven}, \citenamefont {Throssell}, \citenamefont
  {Montgomery}, \citenamefont {Peralta}, \citenamefont {Ogliaro}, \citenamefont
  {Bearpark}, \citenamefont {Heyd}, \citenamefont {Brothers}, \citenamefont
  {Kudin}, \citenamefont {Staroverov}, \citenamefont {Keith}, \citenamefont
  {Kobayashi}, \citenamefont {Normand}, \citenamefont {Raghavachari},
  \citenamefont {Rendell}, \citenamefont {Burant}, \citenamefont {Iyengar},
  \citenamefont {Tomasi}, \citenamefont {Cossi}, \citenamefont {Millam},
  \citenamefont {Klene}, \citenamefont {Adamo}, \citenamefont {Cammi},
  \citenamefont {Ochterski}, \citenamefont {Martin}, \citenamefont {Morokuma},
  \citenamefont {Farkas}, \citenamefont {Foresman},\ and\ \citenamefont
  {Fox.}}]{gaussian09}%
  \BibitemOpen
  \bibfield  {author} {\bibinfo {author} {\bibfnamefont {M.~J.}\ \bibnamefont
  {Frisch}}, \bibinfo {author} {\bibfnamefont {G.~W.}\ \bibnamefont {Trucks}},
  \bibinfo {author} {\bibfnamefont {H.~B.}\ \bibnamefont {Schlegel}}, \bibinfo
  {author} {\bibfnamefont {G.~E.}\ \bibnamefont {Scuseria}}, \bibinfo {author}
  {\bibfnamefont {M.~A.}\ \bibnamefont {Robb}}, \bibinfo {author}
  {\bibfnamefont {J.~R.}\ \bibnamefont {Cheeseman}}, \bibinfo {author}
  {\bibfnamefont {G.}~\bibnamefont {Scalmani}}, \bibinfo {author}
  {\bibfnamefont {V.}~\bibnamefont {Barone}}, \bibinfo {author} {\bibfnamefont
  {G.~A.}\ \bibnamefont {Petersson}}, \bibinfo {author} {\bibfnamefont
  {H.}~\bibnamefont {Nakatsuji}}, \bibinfo {author} {\bibfnamefont
  {X.}~\bibnamefont {Li}}, \bibinfo {author} {\bibfnamefont {M.}~\bibnamefont
  {Caricato}}, \bibinfo {author} {\bibfnamefont {A.}~\bibnamefont {Marenich}},
  \bibinfo {author} {\bibfnamefont {J.}~\bibnamefont {Bloino}}, \bibinfo
  {author} {\bibfnamefont {B.~G.}\ \bibnamefont {Janesko}}, \bibinfo {author}
  {\bibfnamefont {R.}~\bibnamefont {Gomperts}}, \bibinfo {author}
  {\bibfnamefont {B.}~\bibnamefont {Mennucci}}, \bibinfo {author}
  {\bibfnamefont {H.~P.}\ \bibnamefont {Hratchian}}, \bibinfo {author}
  {\bibfnamefont {J.~V.}\ \bibnamefont {Ortiz}}, \bibinfo {author}
  {\bibfnamefont {A.~F.}\ \bibnamefont {Izmaylov}}, \bibinfo {author}
  {\bibfnamefont {J.~L.}\ \bibnamefont {Sonnenberg}}, \bibinfo {author}
  {\bibfnamefont {D.}~\bibnamefont {Williams-Young}}, \bibinfo {author}
  {\bibfnamefont {F.}~\bibnamefont {Ding}}, \bibinfo {author} {\bibfnamefont
  {F.}~\bibnamefont {Lipparini}}, \bibinfo {author} {\bibfnamefont
  {F.}~\bibnamefont {Egidi}}, \bibinfo {author} {\bibfnamefont
  {J.}~\bibnamefont {Goings}}, \bibinfo {author} {\bibfnamefont
  {B.}~\bibnamefont {Peng}}, \bibinfo {author} {\bibfnamefont {A.}~\bibnamefont
  {Petrone}}, \bibinfo {author} {\bibfnamefont {T.}~\bibnamefont {Henderson}},
  \bibinfo {author} {\bibfnamefont {D.}~\bibnamefont {Ranasinghe}}, \bibinfo
  {author} {\bibfnamefont {V.~G.}\ \bibnamefont {Zakrzewski}}, \bibinfo
  {author} {\bibfnamefont {J.}~\bibnamefont {Gao}}, \bibinfo {author}
  {\bibfnamefont {N.}~\bibnamefont {Rega}}, \bibinfo {author} {\bibfnamefont
  {G.}~\bibnamefont {Zheng}}, \bibinfo {author} {\bibfnamefont
  {W.}~\bibnamefont {Liang}}, \bibinfo {author} {\bibfnamefont
  {M.}~\bibnamefont {Hada}}, \bibinfo {author} {\bibfnamefont {M.}~\bibnamefont
  {Ehara}}, \bibinfo {author} {\bibfnamefont {K.}~\bibnamefont {Toyota}},
  \bibinfo {author} {\bibfnamefont {R.}~\bibnamefont {Fukuda}}, \bibinfo
  {author} {\bibfnamefont {J.}~\bibnamefont {Hasegawa}}, \bibinfo {author}
  {\bibfnamefont {M.}~\bibnamefont {Ishida}}, \bibinfo {author} {\bibfnamefont
  {T.}~\bibnamefont {Nakajima}}, \bibinfo {author} {\bibfnamefont
  {Y.}~\bibnamefont {Honda}}, \bibinfo {author} {\bibfnamefont
  {O.}~\bibnamefont {Kitao}}, \bibinfo {author} {\bibfnamefont
  {H.}~\bibnamefont {Nakai}}, \bibinfo {author} {\bibfnamefont
  {T.}~\bibnamefont {Vreven}}, \bibinfo {author} {\bibfnamefont
  {K.}~\bibnamefont {Throssell}}, \bibinfo {author} {\bibfnamefont {J.~A.}\
  \bibnamefont {Montgomery}}, \bibinfo {author} {\bibfnamefont {J.~J.~E.}\
  \bibnamefont {Peralta}}, \bibinfo {author} {\bibfnamefont {F.}~\bibnamefont
  {Ogliaro}}, \bibinfo {author} {\bibfnamefont {M.}~\bibnamefont {Bearpark}},
  \bibinfo {author} {\bibfnamefont {J.~J.}\ \bibnamefont {Heyd}}, \bibinfo
  {author} {\bibfnamefont {E.}~\bibnamefont {Brothers}}, \bibinfo {author}
  {\bibfnamefont {K.~N.}\ \bibnamefont {Kudin}}, \bibinfo {author}
  {\bibfnamefont {V.~N.}\ \bibnamefont {Staroverov}}, \bibinfo {author}
  {\bibfnamefont {T.}~\bibnamefont {Keith}}, \bibinfo {author} {\bibfnamefont
  {R.}~\bibnamefont {Kobayashi}}, \bibinfo {author} {\bibfnamefont
  {J.}~\bibnamefont {Normand}}, \bibinfo {author} {\bibfnamefont
  {K.}~\bibnamefont {Raghavachari}}, \bibinfo {author} {\bibfnamefont
  {A.}~\bibnamefont {Rendell}}, \bibinfo {author} {\bibfnamefont {J.~C.}\
  \bibnamefont {Burant}}, \bibinfo {author} {\bibfnamefont {S.~S.}\
  \bibnamefont {Iyengar}}, \bibinfo {author} {\bibfnamefont {J.}~\bibnamefont
  {Tomasi}}, \bibinfo {author} {\bibfnamefont {M.}~\bibnamefont {Cossi}},
  \bibinfo {author} {\bibfnamefont {J.~M.}\ \bibnamefont {Millam}}, \bibinfo
  {author} {\bibfnamefont {M.}~\bibnamefont {Klene}}, \bibinfo {author}
  {\bibfnamefont {C.}~\bibnamefont {Adamo}}, \bibinfo {author} {\bibfnamefont
  {R.}~\bibnamefont {Cammi}}, \bibinfo {author} {\bibfnamefont {J.~W.}\
  \bibnamefont {Ochterski}}, \bibinfo {author} {\bibfnamefont {R.~L.}\
  \bibnamefont {Martin}}, \bibinfo {author} {\bibfnamefont {K.}~\bibnamefont
  {Morokuma}}, \bibinfo {author} {\bibfnamefont {O.}~\bibnamefont {Farkas}},
  \bibinfo {author} {\bibfnamefont {J.~B.}\ \bibnamefont {Foresman}}, \ and\
  \bibinfo {author} {\bibfnamefont {D.~J.}\ \bibnamefont {Fox.}},\ }\href@noop
  {} {\enquote {\bibinfo {title} {Gaussian09, revision a.02; gaussian, inc.,
  wallingford ct},}\ } (\bibinfo {year} {2016})\BibitemShut {NoStop}%
\bibitem [{\citenamefont {Stanton}\ \emph {et~al.}()\citenamefont {Stanton},
  \citenamefont {Gauss}, \citenamefont {Cheng}, \citenamefont {Harding},
  \citenamefont {Matthews},\ and\ \citenamefont {Szalay}}]{cfour}%
  \BibitemOpen
  \bibfield  {author} {\bibinfo {author} {\bibfnamefont {J.~F.}\ \bibnamefont
  {Stanton}}, \bibinfo {author} {\bibfnamefont {J.}~\bibnamefont {Gauss}},
  \bibinfo {author} {\bibfnamefont {L.}~\bibnamefont {Cheng}}, \bibinfo
  {author} {\bibfnamefont {M.~E.}\ \bibnamefont {Harding}}, \bibinfo {author}
  {\bibfnamefont {D.~A.}\ \bibnamefont {Matthews}}, \ and\ \bibinfo {author}
  {\bibfnamefont {P.~G.}\ \bibnamefont {Szalay}},\ }\href@noop {} {\enquote
  {\bibinfo {title} {{CFOUR, Coupled-Cluster techniques for Computational
  Chemistry, a quantum-chemical program package}},}\ }\bibinfo {note} {{W}ith
  contributions from {A}.{A}. {A}uer, {R}.{J}. {B}artlett, {U}. {B}enedikt,
  {C}. {B}erger, {D}.{E}. {B}ernholdt, {S.} {B}laschke, {Y}. {J}. {B}omble,
  {S.} {B}urger, {O}. {C}hristiansen, {D.} Datta, {F}. Engel, {R}. Faber, {J.}
  {G}reiner, {M}. {H}eckert, {O}. {H}eun, {M}. Hilgenberg, {C}. {H}uber,
  {T}.-{C}. {J}agau, {D}. {J}onsson, {J}. {J}us{\'e}lius, {T}. Kirsch, {K}.
  {K}lein, {G}.{M.} Kopper{W}.{J}. {L}auderdale, {F}. {L}ipparini, {T}.
  {M}etzroth, {L}.{A}. {M}{\"u}ck, {D}.{P}. {O}'{N}eill, {T.} {N}ottoli,
  {D}.{R}. {P}rice, {E}. {P}rochnow, {C}. {P}uzzarini, {K}. {R}uud, {F}.
  {S}chiffmann, {W}. {S}chwalbach, {C}. {S}immons, {S}. {S}topkowicz, {A}.
  {T}ajti, {J}. {V}{\'a}zquez, {F}. {W}ang, {J}.{D}. {W}atts and the integral
  packages {MOLECULE} ({J}. {A}lml{\"o}f and {P}.{R}. {T}aylor), {PROPS}
  ({P}.{R}. {T}aylor), {ABACUS} ({T}. {H}elgaker, {H}.{J}. {A}a. {J}ensen, {P}.
  {J}{\o}rgensen, and {J}. {O}lsen), and {ECP} routines by {A}. {V}. {M}itin
  and {C}. van {W}{\"u}llen. {F}or the current version, see
  http://www.cfour.de.}\BibitemShut {Stop}%
\bibitem [{\citenamefont {Aquilante}\ \emph {et~al.}(2016)\citenamefont
  {Aquilante}, \citenamefont {Autschbach}, \citenamefont {Carlson},
  \citenamefont {Chibotaru}, \citenamefont {Delcey}, \citenamefont {De~Vico},
  \citenamefont {Fdez.~Galv\'{a}n}, \citenamefont {Ferr\'{e}}, \citenamefont
  {Frutos}, \citenamefont {Gagliardi}, \citenamefont {Garavelli}, \citenamefont
  {Giussani}, \citenamefont {Hoyer}, \citenamefont {Li~Manni}, \citenamefont
  {Lischka}, \citenamefont {Ma}, \citenamefont {Malmqvist}, \citenamefont
  {M\"uller}, \citenamefont {Nenov}, \citenamefont {Olivucci}, \citenamefont
  {Pedersen}, \citenamefont {Peng}, \citenamefont {Plasser}, \citenamefont
  {Pritchard}, \citenamefont {Reiher}, \citenamefont {Rivalta}, \citenamefont
  {Schapiro}, \citenamefont {Segarra-Mart\'{i}}, \citenamefont {Stenrup},
  \citenamefont {Truhlar}, \citenamefont {Ungur}, \citenamefont {Valentini},
  \citenamefont {Vancoillie}, \citenamefont {Veryazov}, \citenamefont
  {Vysotskiy}, \citenamefont {Weingart}, \citenamefont {Zapata},\ and\
  \citenamefont {Lindh}}]{molcas8}%
  \BibitemOpen
  \bibfield  {author} {\bibinfo {author} {\bibfnamefont {F.}~\bibnamefont
  {Aquilante}}, \bibinfo {author} {\bibfnamefont {J.}~\bibnamefont
  {Autschbach}}, \bibinfo {author} {\bibfnamefont {R.~K.}\ \bibnamefont
  {Carlson}}, \bibinfo {author} {\bibfnamefont {L.~F.}\ \bibnamefont
  {Chibotaru}}, \bibinfo {author} {\bibfnamefont {M.~l.~G.}\ \bibnamefont
  {Delcey}}, \bibinfo {author} {\bibfnamefont {L.}~\bibnamefont {De~Vico}},
  \bibinfo {author} {\bibfnamefont {I.}~\bibnamefont {Fdez.~Galv\'{a}n}},
  \bibinfo {author} {\bibfnamefont {N.}~\bibnamefont {Ferr\'{e}}}, \bibinfo
  {author} {\bibfnamefont {L.~M.}\ \bibnamefont {Frutos}}, \bibinfo {author}
  {\bibfnamefont {L.}~\bibnamefont {Gagliardi}}, \bibinfo {author}
  {\bibfnamefont {M.}~\bibnamefont {Garavelli}}, \bibinfo {author}
  {\bibfnamefont {A.}~\bibnamefont {Giussani}}, \bibinfo {author}
  {\bibfnamefont {C.~E.}\ \bibnamefont {Hoyer}}, \bibinfo {author}
  {\bibfnamefont {G.}~\bibnamefont {Li~Manni}}, \bibinfo {author}
  {\bibfnamefont {H.}~\bibnamefont {Lischka}}, \bibinfo {author} {\bibfnamefont
  {D.}~\bibnamefont {Ma}}, \bibinfo {author} {\bibfnamefont {P.~r.}\
  \bibnamefont {Malmqvist}}, \bibinfo {author} {\bibfnamefont {T.}~\bibnamefont
  {M\"uller}}, \bibinfo {author} {\bibfnamefont {A.}~\bibnamefont {Nenov}},
  \bibinfo {author} {\bibfnamefont {M.}~\bibnamefont {Olivucci}}, \bibinfo
  {author} {\bibfnamefont {T.~B.}\ \bibnamefont {Pedersen}}, \bibinfo {author}
  {\bibfnamefont {D.}~\bibnamefont {Peng}}, \bibinfo {author} {\bibfnamefont
  {F.}~\bibnamefont {Plasser}}, \bibinfo {author} {\bibfnamefont
  {B.}~\bibnamefont {Pritchard}}, \bibinfo {author} {\bibfnamefont
  {M.}~\bibnamefont {Reiher}}, \bibinfo {author} {\bibfnamefont
  {I.}~\bibnamefont {Rivalta}}, \bibinfo {author} {\bibfnamefont
  {I.}~\bibnamefont {Schapiro}}, \bibinfo {author} {\bibfnamefont
  {J.}~\bibnamefont {Segarra-Mart\'{i}}}, \bibinfo {author} {\bibfnamefont
  {M.}~\bibnamefont {Stenrup}}, \bibinfo {author} {\bibfnamefont {D.~G.}\
  \bibnamefont {Truhlar}}, \bibinfo {author} {\bibfnamefont {L.}~\bibnamefont
  {Ungur}}, \bibinfo {author} {\bibfnamefont {A.}~\bibnamefont {Valentini}},
  \bibinfo {author} {\bibfnamefont {S.}~\bibnamefont {Vancoillie}}, \bibinfo
  {author} {\bibfnamefont {V.}~\bibnamefont {Veryazov}}, \bibinfo {author}
  {\bibfnamefont {V.~P.}\ \bibnamefont {Vysotskiy}}, \bibinfo {author}
  {\bibfnamefont {O.}~\bibnamefont {Weingart}}, \bibinfo {author}
  {\bibfnamefont {F.}~\bibnamefont {Zapata}}, \ and\ \bibinfo {author}
  {\bibfnamefont {R.}~\bibnamefont {Lindh}},\ }\href {\doibase
  https://doi.org/10.1002/jcc.24221} {\bibfield  {journal} {\bibinfo  {journal}
  {J. Comput. Chem.}\ }\textbf {\bibinfo {volume} {37}},\ \bibinfo {pages}
  {506} (\bibinfo {year} {2016})}\BibitemShut {NoStop}%
\bibitem [{mem()}]{memCC}%
  \BibitemOpen
  \href@noop {} {}\bibinfo {note} {The CC calculations performed with CFour
  code are run on a dual socket AMD Epyc 7402 with 256GB of
  memory.}\BibitemShut {Stop}%
\bibitem [{\citenamefont {V{\'{e}}ril}\ \emph {et~al.}(2021)\citenamefont
  {V{\'{e}}ril}, \citenamefont {Scemama}, \citenamefont {Caffarel},
  \citenamefont {Lipparini}, \citenamefont {Boggio-Pasqua}, \citenamefont
  {Jacquemin},\ and\ \citenamefont {Loos}}]{veril_2021}%
  \BibitemOpen
  \bibfield  {author} {\bibinfo {author} {\bibfnamefont {M.}~\bibnamefont
  {V{\'{e}}ril}}, \bibinfo {author} {\bibfnamefont {A.}~\bibnamefont
  {Scemama}}, \bibinfo {author} {\bibfnamefont {M.}~\bibnamefont {Caffarel}},
  \bibinfo {author} {\bibfnamefont {F.}~\bibnamefont {Lipparini}}, \bibinfo
  {author} {\bibfnamefont {M.}~\bibnamefont {Boggio-Pasqua}}, \bibinfo {author}
  {\bibfnamefont {D.}~\bibnamefont {Jacquemin}}, \ and\ \bibinfo {author}
  {\bibfnamefont {P.-F.}\ \bibnamefont {Loos}},\ }\href {\doibase
  10.1002/wcms.1517} {\bibfield  {journal} {\bibinfo  {journal} {WIREs Comput.
  Mol. Sci.}\ }\textbf {\bibinfo {volume} {n/a}},\ \bibinfo {pages} {e1517}
  (\bibinfo {year} {2021})}\BibitemShut {NoStop}%
\end{thebibliography}%

\end{document}